\documentclass[10pt,twocolumn]{article}

\usepackage{graphicx}
\usepackage{epstopdf}

\usepackage{cite}

\usepackage{amsmath}
\usepackage{amssymb}
\usepackage{amsfonts}

\usepackage[dvipdfm,colorlinks,urlcolor=blue,linkcolor=blue,citecolor=blue,anchorcolor=blue]{hyperref}

 \usepackage[margin=0.5in]{geometry} 

\begin{document}

\title{Multicharged optical vortices induced in a dissipative atomic vapor system}

\author{Yiqi Zhang,$^{1,*}$  Milivoj R. Beli\'c,$^{2}$ Zhenkun Wu,$^1$ Chenzhi Yuan,$^1$ \\
Ruimin Wang,$^1$ Keqing Lu,$^3$ and Yanpeng Zhang$^{1,**}$ \\
$^1$Key Laboratory for Physical Electronics and Devices
of the Ministry of Education \& \\
Shaanxi Key Lab of Information Photonic Technique,\\
Xi'an Jiaotong University, Xi'an 710049, China \\
$^2$Science Program, Texas A\&M University at Qatar, P.O. Box 23874 Doha, Qatar \\
$^3$School of Information and Communication Engineering, \\
Tianjin Polytechnic University, Tianjin 300160, China \\
$^*$zhangyiqi@mail.xjtu.edu.cn, \quad
$^{**}$ypzhang@mail.xjtu.edu.cn
}
\date{}
\maketitle

\begin{abstract}
We investigate numerically the dynamics of optical vortex beams carrying different topological charges,
  launched in a dissipative three-level ladder-type nonlinear atomic vapor.
  We impose the electromagnetically induced transparency (EIT) condition on the medium.
  Linear, cubic, and quintic susceptibilities, considered simultaneously with the dressing effect,
  are included in the analysis.
  Generally, the beams slowly expand during propagation and new vortices are induced,
  commonly appearing in oppositely-charged pairs.
  We demonstrate that not only the form and the topological charge of the incident beam,
  but also its growing size in the medium greatly affect the formation and evolution of vortices.
  We formulate common rules for finding the number of induced vortices and the corresponding rotation directions,
  stemming from the initial conditions of various incident beams, as well as from the dynamical aspects of their propagation.
  The net topological charge of the vortex is conserved during propagation, as it should be,
  but the total number of charges is not necessarily same as the initial number,
  because of the complex nature of the system.
  When the EIT condition is lifted, an enhancement region of beam dynamics is reached,
  in which the dynamics and the expansion of the beam greatly accelerate.
  In the end, we discuss the liquid-like behavior of light evolution in this dissipative system
  and propose a potential experimental scheme for observing such a behavior.
\end{abstract}

\section{Introduction}

Vortices are physical objects commonly observed in nature.
They still consume a lot of interest and research, both in physical and engineering
sciences.
A vortex possesses a phase singularity and an energy flow around this
singularity, rotating clockwise or counterclockwise.
Optical vortices have also become a hot topic in recent decades,
for their potential applications in many fields \cite{tikhonenko_josab_95, crabtree_ao_2004, grier_nature_2003, kuga_prl_1997, mair_nature_2001, gorodetski_prl_2013}.
To date, research on optical vortices in different materials, including
bulk media \cite{kivshar_2005},
discrete systems \cite{lederer_pr_2008},
atomic vapors \cite{skupin_prl_2007,wang_oe_2012},
dissipative optical systems \cite{mihalache_prl_2006},
and Bose-Einstein condensates \cite{theocharis_prl_2003},
has been reported.

In a dissipative system based on the model of laser cavities \cite{he_oe_2007},
the multicharged necklace-shaped beams merge into stable fundamental or vortex solitons.
The two-dimensional (2D) complex Ginzburg-Landau (CGL) equation is found to adequately describe this model.
In Refs. \cite{michinel_prl_2006,michinel_pre_2002},
multicharged multidimensional solitons and light condensates were found via atomic coherence
that appear analogous to the usual liquids. The liquid-like beam has a flat top with sharp decaying edge.
In fact, flat top solitons obeying CGL were reported earlier \cite{saarloos_prl_1990, saarloos_pd_1992}.
Dissipative solitons of CGL with cubic-quintic (CQ) nonlinearity were also studied before
\cite{ankiewicz_pra_2008, chang_josab_2008, grelu_josab_2010}.
In addition, studies on the evolution of multicharged vortices in second-harmonic generation
revealed that the topological charge, beam width, light intensity, and evolution distance may
affect the number of induced vortices \cite{molina_josab_2000}.

In this paper we investigate the evolution of
multicharged necklace-shaped and azimuthon-shaped beams in multilevel atomic vapors
with giant third-order and fifth-order nonlinear susceptibilities of opposite signs.
The system is dissipative and placed in a laser cavity. A generic laser beam is incident on
the medium, that can assume different forms, depending on the values of different beam parameters.
Its evolution is followed by solving a Ginzburg-Landau-type CQ nonlinear Schr\"odinger
equation (NLSE) that describes well the behavior of the slowly-varying amplitude of the optical electric field.
Various cases of beam incidence are discussed in some detail. We did not locate any stable beam
propagation, although such incidences should exist -- we were simply not looking for them. Our attention was confined
to studying the dynamics of slowly varying beams in propagation over considerable, experimentally attainable distances.
Thus, our study was more geared toward direct comparison with experiment.

The paper is organized as follows: In Sec. \ref{modelling} we briefly introduce the general model equation
for beam propagation in dissipative atomic vapors and in Sec. \ref{vortex_part} we consider the
simple vortex and necklace-shaped beam incidence on the medium. We then discuss in detail the propagation properties
of azimuthon-shaped beams in Sec. \ref{azimuthon_part}.
In Sec. \ref{enhancement} we cover the dynamics of the system in the enhancement region.
In Sec. \ref{droplet}, liquid-like behavior of light evolution in our system and a potential experimental method to
investigate this behavior are described.
Section \ref{conclusion} concludes the paper.

\section{Mathematical modeling}
\label{modelling}

\begin{figure}[htbp]
\centering
\includegraphics[width=0.3\columnwidth]{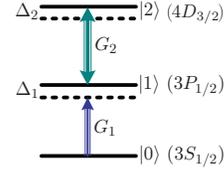}
\caption{(color online) Schematics of energy levels.
}
\label{fig1}
\end{figure}

We consider paraxial propagation of a probe vortex beam
in a ladder-type three-level atomic system, formed by the $3S_{1/2}$, $3P_{1/2}$ and $4D_{3/2}$ levels of sodium,
as shown in Fig. \ref{fig1}.
The propagation model is based on the paraxial wave equation of the form \cite{michinel_prl_2006,mihalache_prl_2006}:

\begin{equation}\label{initial_eq}
i\partial _z E_1  + \frac{1-i\beta}{{2k}}\nabla^2_\bot E_1  + \frac{k}{2}\chi E_1  = 0,
\end{equation}
for the slowly-varying amplitude of the electric field $E_1$ of the beam.
Here $\nabla^2_\bot=\partial_{xx}+\partial_{yy}$ is the transverse Laplacian,
$k$ is the wave number, $\chi$ is the total susceptibility of the atomic vapor system,
and $\beta$ is the diffusion coefficient, which originates from the model of laser cavity.
It has been demonstrated before that vortices (even vortex solitons) can be supported in the framework of a 2D CQ CGL model with the $\beta$ term present \cite{skryabin_prl_2002,mihlache_pra_2007}.

The involved linear, third-order and fifth-order nonlinear susceptibilities in Eq. (\ref{initial_eq}) are:

\begin{align}
\label{chis1}  \chi^{(1)} = &\frac{\eta}{K},\\
\label{chis3}  \chi^{(3)}|E|^2 = &\frac{\eta}{K^2}\left(\frac{|G_1|^2}{d_1'}+\frac{G_2^2}{d_2}\right),\\
\label{chis5}  \chi^{(5)}|E|^4 = &\frac{\eta}{K^3}\left(\frac{|{G_1}|^4}{d'_1{}^2}+\frac{G_2^4}{d_2^2}+\frac{|G_1|^2 G_2^2}{d'_1d_2}\right),
\end{align}
with $\eta=iN\mu _{10}^2/(\hbar \epsilon _0)$, $K = d_1 + G_2^2 /d_2$,
$d_1=\Gamma _{10}  + i\Delta _1$,
$d'_1=\Gamma _{11}  + G_2^2/(\Gamma_{21}+i\Delta _2)$, and
$d_2=\Gamma _{20}  + i\left( {\Delta _1  + \Delta _2 } \right)$.
$G_1=\mu_{10} E_1/\hbar$ is the Rabi frequency of the probe field,
$G_2$ is the Rabi frequency of the couple field, and
$\Delta_{1,2}$ are the detunings of the probe and couple fields.
Further, $N$ is the atomic density,
$\Gamma_{ij}$ denotes the population decay rate between the corresponding energy levels $|i\rangle$ and $|j\rangle$,
and $\mu_{10}$ is the electric dipole moment.
Combining Eqs. (\ref{chis1})-(\ref{chis5}), we obtain the total susceptibility

\begin{equation}\label{chis}
  \chi = \chi^{(1)} + \chi^{(3)}|E|^2 + \chi^{(5)}|E|^4.
\end{equation}
In Appendices, we present briefly the derivation of this susceptibility.

Thus, the propagation equation appears to be of the CQ NLSE-type, which
seems to be an appropriate general model for the description of the interaction
of an atomic vapor system with laser radiation.
The second and the third term in $\chi$ represent the cubic and the quintic contributions to the total susceptibility.
Furthermore, it is noted that Eq. (\ref{initial_eq}) is an equation of the CGL form,
with the diffusion coefficient $\beta$ coming from the model of the laser cavity.
The presence of $\beta$ is crucial in the modeling, but the actual value, which we take to be of the order of 1,
when varied does not change much the beam distribution, as it will be seen below.

We launch in this atomic system incident beams of generic form \cite{zhang_ol_2012}:

\begin{equation}\label{azimuthon}
\begin{split}
      \psi \left( {z=0,r,\theta } \right) = & A{\rm{sech}}\left[ \left(r - R_0\right)/r_0 \right] \times \\
          &[\cos(n\theta)+iB\sin(n\theta)] e^{il\theta},
\end{split}
\end{equation}
where $A$ is the amplitude, $R_0$ the mean radius, $r_0$ the width of the beam,
$1-B$ the modulation depth,
$l$ the input topological charge, and $2n$ the number of necklace beads in the input beam.
The choice of $B$ and $n$ parameters governs the shape of the beam. Thus, for $B=1$ or $n=0$
one obtains a simple vortex, for
$B=0,~n\neq 0$ a necklace, and for
$B\neq0,~n\neq0$ an azimuthon \cite{zhang_oe_2010}.

From the mathematical point of view,
the beam in Eq. (\ref{azimuthon}) can be viewed as a superposition of two vortices, i.e.,

\begin{align}\label{splitvortex} 
 \psi \left( {z=0,r,\theta } \right) =&  {\rm{sech}}\left( \frac{r - R_0}{r_0} \right) \left[\frac{A}{2}(1+B)e^{i(l+n)\theta} + \right. \notag \\
 & \left. \frac{A}{2}(1-B)e^{i(l-n)\theta} \right].
\end{align}

The net topological charge (NTC) of a vortex beam is defined as  \cite{kivshar_2005, molina_josab_2000}

\begin{equation}
  {\rm{NTC}}=\frac{1}{2\pi}\oint \nabla\Phi \cdot d\textbf{l},
\end{equation}
where $\Phi$ is the total beam phase and $d\textbf{l}$ the line element of a closed path that surrounds all the singularities.
Hence, the NTCs of the component vortices given in Eq. (\ref{splitvortex}) are $l+n$ and $l-n$, respectively.
However, what NTCs will be observed upon propagation in this nonlinear system is not so simply determined.

When $B \neq 0$ and in light of the fact that $A(1+B)/2>A(1-B)/2>0$,
the initial NTC of the beam in Eq. (\ref{splitvortex}) is $l+n$ \cite{molina_josab_2000}.
If $B=0$, the vortex with a larger magnitude of NTC in Eq. (\ref{splitvortex}) will dissipate faster during propagation,
due to the diffusion term in Eq. (\ref{initial_eq}) \cite{he_oe_2007}.
Thus, the vortex with a smaller NTC $\min\{|l\pm n|\}$ will remain stable for longer during propagation
and the overall NTC will correspond to that charge, instead of the one directly calculated from Eq. (\ref{azimuthon}).
To clarify, concerning the overall NTC, we end up with

\begin{equation}\label{NTCs2}
  \textrm{NTC}=\left\{
  \begin{split}
  &l+n,& B\neq 0 \\
  &\sigma\min\{|l\pm n|\},& B=0
  \end{split}
  \right.
\end{equation}
for the beam in Eqs. (\ref{azimuthon}) and (\ref{splitvortex}).
Here the sign $\sigma=\pm1$ is the same as the sign of the NTC with a smaller absolute value.

In accordance with previous literature \cite{phillips_prl_2001, lukin_nature_2001, michinel_prl_2006, wu_josab_2008,steck_sodium},
we set the parameters to:
$N=10^{13}~{\rm{cm}}^{-3}$,
$\mu_{10}=3\times 10^{-29}~{\rm{C\cdot m}}$,
$\Delta_1=1~{{\rm{MHz}}}$,
$\Delta_2=-1~{{\rm{MHz}}}$,
$G_2=40~{{\rm{MHz}}}$,
$\Gamma_{10}=2\pi\times4.86~{{\rm{MHz}}}$,
$\Gamma_{20}=2\pi\times0.485~{{\rm{MHz}}}$,
$\Gamma_{21}=2\pi\times6.36~\rm MHz$,
$\Gamma_{11}=2\pi\times5.88~\rm MHz$,
$\lambda=600~{\rm{nm}}$,
$r_0=100~\mu{\rm{m}}$,
$R_0=200~\mu{\rm{m}}$.
As it is well known, $\Delta_1+\Delta_2=0$ corresponds to a two-photon resonance condition,
which brings in the electromagnetically induced transparency (EIT) \cite{harris_pt_1997,fleischhauer_rmp_2005}.
When $\Delta_1+\Delta_2\neq0$, the system is moved to the enhancement region \cite{zhang_ieee_2012}.

\section{Simple vortex and necklace incidence}
\label{vortex_part}

\begin{figure}[htbp]
  \centering
  {\includegraphics[width=\columnwidth]{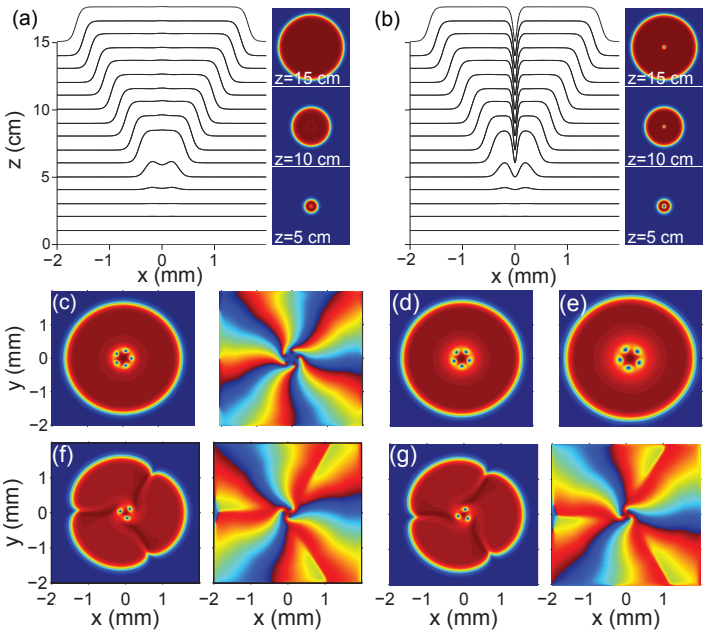}}
  \caption{(color online) (a) and (b) Evolution of vortex incidences during propagation for
  $l=0$ and $1$, respectively. The right panels are selected intensities when the beam propagates to
  $z=5,~10,~\textrm{and}~15~\rm{cm}$, respectively.
  (c) Intensities and phases of vortices
  with $l=5$ at $z=15~\rm cm$. In (a)-(c) $\beta=0.5$.
  (d) and (e) Intensities of vortices with $l=5$ at $z=15~\rm cm$ when $\beta=1$ and $\beta=2$, respectively.
  (f) and (g) Intensities and phases of tri-lobe inputs ($n=1.5$) at $z=15~\rm cm$
  corresponding to $l=-4.5$ and $4.5$, respectively.
  }
  \label{fig2}
\end{figure}

Evolution of a simple fundamental vortex with $l=0,~n=0$ is depicted in Fig. \ref{fig2}(a).
Upon propagation, the input notch of the vortex at the origin disappears and
the vortex changes into a super-Gaussian-like pulse
whose width and amplitude grow fast.
A saturable plateau is reached at $\sim 10~{\rm{V/m}}$ \cite{zhang_ol_2012},
as seen from the cross sections of intensity profiles at $y=0$ upon propagation in Figs. \ref{fig2}(a) and (b);
these results are similar to the results presented in Refs. \cite{kivshar_2005,michinel_prl_2006}.
On the other hand, if we consider evolution of a simply-charged vortex ($l=1$),
the width and the amplitude grow similarly
but the singularity at the origin remains, as shown in Fig. \ref{fig2}(b).

The saturation phenomenon originates from the competition
within the CQ nonlinearity in the atomic system.
The nonlinearity is defocusing and capable of supporting stable vortices \cite{kivshar_2005}.
However, the beam widens during propagation, because the nonlinearity is too weak to balance diffraction and form
a stable soliton. If $l$ is increased to 5, the corresponding number of singularities will form around the origin,
as shown in Fig. \ref{fig2}(c).
From the corresponding phase gradients, we infer that every notch is indeed a vortex.

In Figs. \ref{fig2}(a)-(c), we did set $\beta=0.5$;
however, the precise value of $\beta$ is not so important for our results.
When the simulation is redone with $\beta=1$ and 2, qualitatively the same results are obtained,
as shown in Figs. \ref{fig2}(d) and (e).
It can be seen from Fig. \ref{fig2}(c) and the results at $z=15~\rm cm$ in Figs. \ref{fig2}(d) and (e) that
the value of $\beta$ does not affect the shape of the beam but the speed of the beam spreading.
So, we fix $\beta=0.5$ in the following investigations.

Concerning necklace beams, in which $2n$ determines the number of input lobes,
we can have a necklace incidence with an even or odd number of beads by taking $n$ an integer
or half-integer, respectively \cite{he_pra_2008}.
Figures \ref{fig2}(f) and (g) display intensity and phase snapshots of
the necklace with $n=1.5$, $l=-4.5$ and 4.5 at $z=15~\rm cm$, respectively.
A comparison of them
reveals that the number of notches that appear in the beam is determined by $\min\{|l\pm n|\}$,
which is also the absolute value of the overall NTC given in Eq. (\ref{NTCs2}).
Even though Figs. \ref{fig2}(f) and (g) show the same number of vortices upon propagation,
they rotate in the opposite directions,
because the corresponding total NTCs are $-3$ and $3$, respectively.


\begin{table*}[htbp]
  \centering
  \caption{Induced vortices for necklace incidence.}
  \begin{tabular}{*{9}{c}} \\ \hline
  & $l < -n$ & $l=-n$ & $-n < l < 0$ & $l=0$ & $0<l<n$ & $l=n$ & $l>n$  \\ \hline
  & $\circlearrowright$ & - & $\circlearrowleft$ & - & $\circlearrowright$ & - & $\circlearrowleft$ \\
  No. & $-n-l$ & 0 & $n+l$ & 0 & $n-l$ & 0 & $-n+l$ \\ \hline
  \end{tabular}
  \label{table1}
\end{table*}

Based on a number of simulations, one can formulate certain rules for calculation
of the number of vortices and for determining their rotation senses.
These rules are presented in Table \ref{table1}.
The symbols $\circlearrowleft$ and $\circlearrowright$ in the table represent rotation senses of the vortices,
which are opposite to the phase gradients.
We can see from Fig. \ref{fig2} and above analysis that the NTC for both the two cases is conserved.

\section{Azimuthon incidence}
\label{azimuthon_part}

Apparently, the azimuthon case, with both $B\neq0,~n\neq0$, is more complex.
Our numerical simulations indicate the existence of two critical values
$r_{cr1}\approx 42.5 ~\mu m$ and $r_{cr2}\approx 36.1 ~\mu m$ for the beam width $r_0$.
The behavior of wide beams, when $r_0>r_{cr1}$, and of the narrow beams, when $r_0<r_{cr2}$,
is relatively simple. The vicinity of the beam edge in the first case and of the beam core
in the second enforce a simple predictable whole-beam behavior. The case in-between, $r_{cr2}<r_0<r_{cr1}$,
is more complicated, as the influences of the core and of the edge compete.
Thus, a varied behavior is observed there.

\subsection{The case $r_0>r_{cr1}$; the wide beam}

\begin{figure}[htbp]
  \centering
  {\includegraphics[width=\columnwidth]{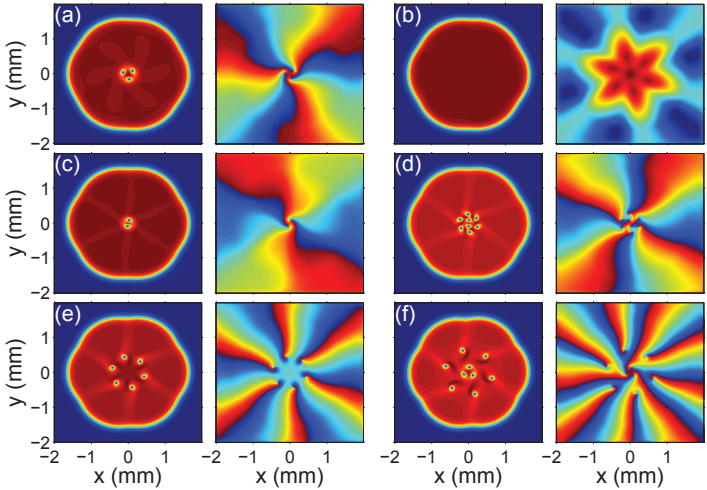}}
  \caption{(color online) Intensities and phases of azimuthons with $n=3$ and $B=0.5$ at $z=15~\rm cm$ for
  $l=-6$ (a), $-3$ (b), $-1$ (c), 1 (d), 3 (e), and 6 (f), respectively.
  }
\label{fig3}
\end{figure}

The intensities and phases at $z=15~\rm cm$ of a hexapole azimuthon in the case $r_0>r_{cr1}$, with $n=3$ and different $l$,
is displayed in Fig. \ref{fig3}.

Similar to the necklace incidence case,
we can formulate the common rules for the azimuthon case,
which are summarized in Table \ref{table2}.
According to these rules,
the number of notches at the origin is determined by NTC;
so, from (a) to (f) in Fig. \ref{fig3},
the number of notches should be 3, 0, 2, 2, 0, and 3, respectively.
However, in (d)-(f),
there are additional 6 notches around the origin, equal to the number of beads.
The explanation for this occurrence is that there is an energy flow around each of the phase singularities,
and at the same time the beads fuse, due to the influence of the diffusion coefficient $\beta$.
Hence, when the speed of energy flow is greater than the fusion speed of the beads,
at the twist of the beads new vortices will form.

From the phases shown in Fig. \ref{fig3},
we can analyze the energy flow direction of the induced vortices,
as well as the overall NTC of the beam during propagation.
Taking Fig. \ref{fig3}(d) as an example,
the phase gradients of the inner 2 and the outer 6 vortices are along opposite directions.
Thus, the NTC is $+1$ for each outer vortex, but $-1$ for the inner vortex.
Considering that the overall NTC of the beam is a sum of all the NTCs,
the NTC for Fig. \ref{fig3}(d) is $4$, which equals $l+n$.
We obtain the same result, that the overall NTC is $l+n$,
if we choose other evolutions for Fig. \ref{fig3}.
The overall NTC in our numerical simulations agrees well with that given in Eq. (\ref{NTCs2}),
and furthermore, it is conserved during propagation.

\begin{table*}[htbp]
  \centering
  \caption{Induced vortices for azimuthon incidence when $r_0>r_{cr1}$.}
  \begin{tabular}{*{9}{c}} \\ \hline
& $l < -n$ & $l=-n$ & $-n < l < 0$ & $l=0$ & $0<l<n$ & $l=n$ & $l>n$  \\ \hline
outer & - & - & - & - & $\circlearrowleft$ & $\circlearrowleft$ & $\circlearrowleft$ \\
inner & $\circlearrowright$ & - & $\circlearrowleft$&  $\circlearrowleft$& $\circlearrowright$ & - & $\circlearrowleft$ \\
No. & $-n-l$ & 0 & $n+l$ & $n$ & $3n-l$ & $2n$ & $n+l$ \\ \hline
  \end{tabular}
  \label{table2}
\end{table*}

\subsection{The in-between case: $r_{cr2}<r_0<r_{cr1}$}\label{azimuthon_two}

\begin{figure}[htbp]
  \centering
  {\includegraphics[width=\columnwidth]{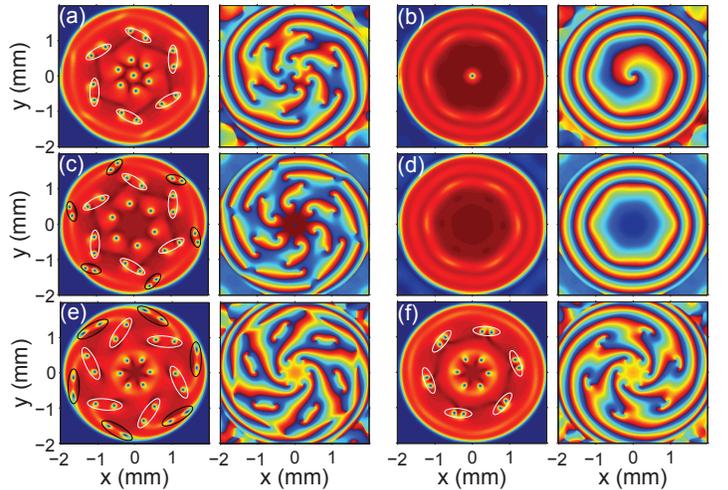}}
  \caption{(color online) Intensities and phases of azimuthons
  with $n=3,~B=0.5$, and $r_0=40~\mu {\rm{m}}$ at $z=30~\rm cm$
  for $l=2$ (a), $-2$ (b), $3$ (c), $-3$ (d), $9$ (e), and $-9$ (f), respectively.
  The vortices placed in white and black ellipses are vortex pairs.
  }
  \label{fig4}
\end{figure}

The evolution of a hexapole azimuthon, with $n=3$ and different $l$
is presented in Figs. \ref{fig4}(a)-(f),
depicting the intensities and phases of different azimuthons at $z=30~\rm cm$.
The beam width is set to
$r_0=40~\mu{\rm{m}}$, which is in-between $r_{cr2}$ and $r_{cr1}$.

According to the rules from Table \ref{table2},
the total number of induced vortices can be predicted.
However, from Fig. \ref{fig4} it is seen that
more and more vortices appear during propagation.
The reason is that the energy flow dictates what happens;
the speed of the energy flow is the largest, as compared to the speed of the fusion and spreading.
From the phases at different propagation distances,
it is visible that the energy flow brought by the vortices around the origin of the fused beam
is always faster than that at the edge of the beam;
this asynchrony preferably forms new phase singularities at the edge of the beam,
and new vortices are induced correspondingly.
Even though one cannot give certain rules for this case, because the number of induced vortices is greatly affected
by the beam width, one can still predict the NTC of the beam.

For the case shown in Fig. \ref{fig4}(a),
the vortices can be divided into three kinds, the inner one, the middle six, and the outer twelve, surrounded by ellipses.
From the corresponding phases
we find that the inner and the middle six vortices share different rotation senses,
so that the total NTC of the seven vortices is five, which can be calculated from $l+n$ given in Eq. (\ref{NTCs2}).
The outer twelve vortices appear as vortex pairs, which can be seen in Fig. \ref{fig4}(a).
The two vortices in the vortex pair rotate in the opposite ways,
i.e., the topological charges of the two vortices are $\pm1$, respectively.
So, the sum NTC of the two vortices is zero,
which does not affect the total topological charge of the beam during propagation.
Thus, the NTC of the beam in Fig. \ref{fig4}(a) is $l+n$ and it is conserved.
Even though the case in Fig. \ref{fig4}(b) should show additional six vortices besides the vortex at the origin during propagation,
these ultimately annihilate, due to the fact that new phase singularities cannot form, because of the fast spread and fusion of the beams;
in any case, the NTC still complies with the rule $l+n$ and is conserved.

The same results are obtained when the cases shown in Figs. \ref{fig4}(c), (d) and (f) are discussed.
Specially in Fig. \ref{fig4}(c), in addition to the 6 vortex pairs surrounded by the white ellipses,
another 6 vortex pairs surrounded by the black ellipses are induced during propagation.
In light of the fact that vortex pairs will not add extra topological charge to the beam,
the NTC for the cases mentioned above is always $l+n$.
And for these cases, regardless of the number of vortex pairs,
Table \ref{table2} can help us determine the total number of vortices and their rotation senses.

Abnormally, the NTC of the case in Fig. \ref{fig4}(e) is $l-n$,
and our numerical simulations demonstrate that the NTC for the cases with $l>n$ is always $l-n$ instead of $l+n$
(other numerical results corroborating this fact are not shown in this paper).
Even though this number is different from the initial total NTC $l+n$ in Eq. (\ref{NTCs2}), it is still conserved during propagation.
All in all, the NTC of the beam is conserved during propagation,
but the final NTC need not be the same as that of the incident beam, because of the dissipative property of the system.
This property enables the diffusion term to annihilate the vortex component with high topological charge.

\subsection{The case $r_0<r_{cr2}$; the narrow beam}

\begin{figure}[htbp]
  \centering
  {\includegraphics[width=0.75\columnwidth]{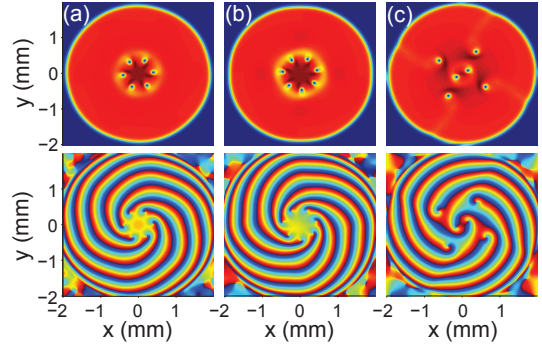}}
  \caption{(color online) Intensities and phases of azimuthons with $r_0=30~\mu {\rm{m}}$ at $z= 30~\rm cm$,
  (a) $l=9$, $n=3$, (b) $l=9$, $n=2$,  and (c) $l=4$, $n=2$, respectively.
  }
\label{fig5}
\end{figure}

In this part, we set $r_0=30~\mu \textrm{m}$, which fulfills the condition $r_0<r_{cr2}$.
According to our numerical simulations,
the case with $l\leq n$ is without vortex pairs appearing during propagation,
and it shares rules given in Table \ref{table2}.
Hence, here we do not show the corresponding results and provide no discussion.
For the $l>n$ case,
we first redo the propagation of the beam used in Fig. \ref{fig4}(e);
we find that the number of induced vortices is six,
which can be calculated from $l-n=6$,
as shown by the intensity and phase in Fig. \ref{fig5}(a).

However, the exact rule for $n=2$ is not certain, because the energy flow is weakened,
and whether it can produce more vortices or not, it is up to the value of $l$ set for the initial beam.
If $l$ is bigger, the fusion will be accelerated,
so the production of new vortices will be limited.
The value $l=5$ appears to be a boundary for this case: the rule is $l-n$ for $l>5$ and $l+n$ for $l\leq5$.
Figures \ref{fig5}(b) and (c) display two numerical simulations corresponding to $l=9$ and $l=4$,
in which the number of vortices are $l-n=7$ and $l+n=6$, respectively. They comply with the rule.

The outer four vortices that appear in Fig. \ref{fig5}(c) are induced from the fusion process of the incident beads,
the number $2n$ of which will determine the number of the outer induced vortices.
So, the total number of vortices is $(l-n)+2n=l+n$.
The evolution of incident azimuthons with $n=1$ and $n>3$ is also simulated,
and we find that $l+n$ and $l-n$ vortices are induced during propagation, respectively.
The lower the number of beads,
the more energy each bead will possess, which will strengthen the energy flow.
That is why the number of induced vortices is $l+n$ if $n=1$.
In other words,
the NTC of the azimuthons with $n=1$ and $n=2,~l\leq 5$ in this part complies with the rule given in Eq. (\ref{NTCs2}).

\begin{table}[htbp]
  \centering
  \caption{Induced vortices for azimuthon incidence when $r_0<r_{cr2}$ and $l>n$.}
  \begin{tabular}{*{9}{c}} \\ \hline
& \multicolumn{2}{c}{$n=1$} & \multicolumn{3}{c}{$n=2~(l\leq5|l>5)$} &\multicolumn{2}{c}{$n\geq3$} \\ \hline
outer & \multicolumn{2}{c}{$\circlearrowleft$} & \multicolumn{3}{c}{$\circlearrowleft|\circlearrowleft$} & \multicolumn{2}{c}{-} \\
inner & \multicolumn{2}{c}{$\circlearrowleft$} & \multicolumn{3}{c}{$\circlearrowleft|\circlearrowleft$} & \multicolumn{2}{c}{$\circlearrowleft$} \\
No. & \multicolumn{2}{c}{$l+n$} & \multicolumn{3}{c}{$l+n|l-n$} & \multicolumn{2}{c}{$l-n$} \\ \hline
  \end{tabular}
  \label{table3}
\end{table}

Based on numerics,
we can develop the common rules for this case, as exhibited in Table \ref{table3}.
Concerning the NTC of the beam, we claim that it is still conserved for each case during propagation
but is not necessarily the same as the incident one; this
is similar to the NTC conservation rule given in part \ref{azimuthon_two}.

\section{The enhancement region}
\label{enhancement}

\begin{figure}[htbp]
  \centering
  {\includegraphics[width=\columnwidth]{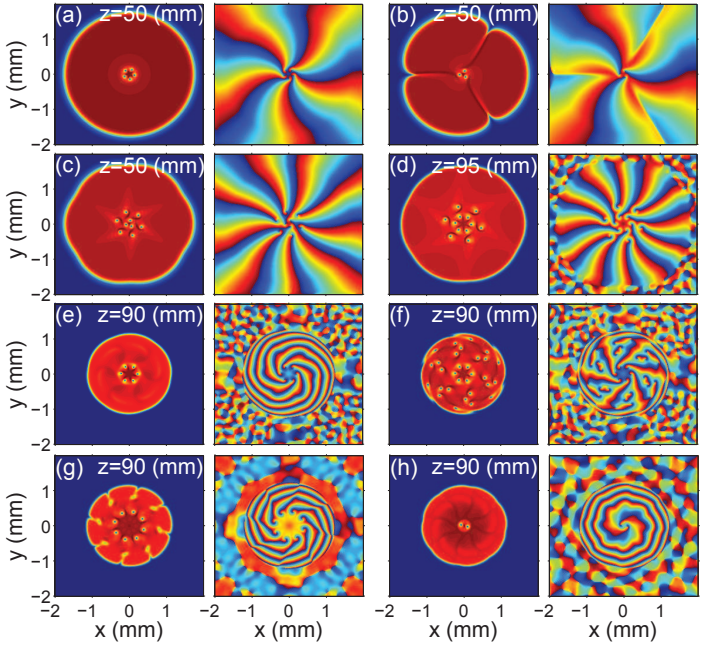}}
  \caption{(color online) Intensities and phases of (a) vortex with $l=5$, $n=0$,
  (b) necklace with $l=4.5$, $n=1.5$, and (c) azimuthon with $l=6$, $n=3$, for $r_0=100~\mu {\rm{m}}$, respectively.
  (d)-(f) Azimuthons with $l=9$, $n=3$, for $r_0=40~\mu {\rm{m}}$, $20~\mu {\rm{m}}$, and
  $23~\mu {\rm{m}}$,  respectively.
  (g) and (h) Azimuthons with $r_0=20~\mu {\rm{m}}$, $n=4$, for $l=4$ and $l=6$, respectively.}
\label{fig6}
\end{figure}

Now we set $\Delta_1=1.1~{\rm MHz}$ and keep $\Delta_2$ fixed at $-1$,
so that $\Delta_1+\Delta_2 \neq 0$ and the EIT condition is lifted.
We do the corresponding simulations and the results are shown in Fig. \ref{fig6}.
Compared with the results obtained under the EIT condition, we see that
the phenomena observed and laws formulated before are still applicable,
but there are several differences.
Firstly, the beams now spread more quickly, so that we can observe almost the same results after only one third of the distance covered before.
Secondly, the saturable plateau shifts to $\sim 60~{\rm V/m}$, which is much higher than before.
Last but not the least, the two threshold values of the azimuthon size $r_0$ become smaller.

We would like to note that the boundary values $n=2$ and $l=5$ in the rule displayed in Table \ref{table3} are not suitable for this case.
The boundary values now are $n=4$ and $l=5$, indicated by more numerical results. An example is shown in Figs. \ref{fig6}(g) and (h),
which are done for $l=4$ and $l=6$, respectively.
The reason for the change is that the beam spreads faster in this case,
so that the outside vortices appear at a shorter propagation distance and have relatively longer distance from the beam edge.
A number of numerical simulations demonstrates that the vortices induced near the boundary of the beam may annihilate,
and so, the higher the value of $n$, the shorter the distance between the outside vortices and the beam edge.
Correspondingly, the boundary value of $n$ for this case should increase.
We also performed a numerical simulation with $\Delta_1=0.9~{\rm MHz}$ (not shown here),
from which we find that the beams also spread but the maximum of the plateau is reduced;
this means that the loss plays a more prominent role now.


\section{The liquid-like behavior of light and potential experiment}
\label{droplet}

Results presented in the above sections point to the development of beam structures with flat tops and sharply decaying edges.
These features are characteristic of the so-called droplets of light that tend to form in a system with competing CQ nonlinearities
\cite{michinel_prl_2006,paz-alonso_pre_2004,paz-alonso_prl_2005,michinel_pre_2002},
in which diffraction, self-focusing nonlinearity, and self-defocusing nonlinearity
are regarded as the kinetic term, the cooling mechanism, and van der Waals (vdW) force, respectively.
Hence, the framework described by Eq. (\ref{initial_eq}) in this paper contains key ingredients to guarantee the formation of light droplets.

The sharp decaying edge illustrates the strong surface tension of the light droplet, due to vdW force,
which drives the formation of a circular flat top, so that the surface tension is uniform everywhere
and helps the pressure on the surface of light droplet reach an equilibrium state \cite{gennes_rmp_1985}.

Thus far, liquid-like beams resulting from the signal and probe beams generated in a four-wave mixing process with atomic coherence
were experimentally observed in sodium atomic vapors under the condition of competition between the third- and fifth-order nonlinear susceptibilities \cite{wu_prl_2013}.
Hence, the liquid-like beams with multi-vortices and vortex pairs investigated in this paper can be possibly observed in a similar experiment,
even though such an experiment would be more complicated.
According to the multi-beam interference technique \cite{zhangyanpeng_oe_2010},
the interference patterns can be used to produce an incident beam with several beads.
The more beams participate in the interference, the more beads the interference pattern will produce \cite{zhangyanpeng_oe_2010}.
Assuming the topological charge in the multi-beam interference pattern is 1,
one can put a phase mask \cite{chen_prl_1997} behind the interference pattern, to add more topological charges to the incident beam,
according to the requirements.
Strong $G_2$ guarantees that the third- and fifth-order nonlinear susceptibilities will play role in this system.
One can change the temperature of the atomic vapor to adjust the atomic density, which
is equivalent to the variation of the evolution distance.

\section{Conclusion}
\label{conclusion}

In conclusion, we have demonstrated that optical vortices can form
from vortex, necklace, and azimuthon incidences with different topological charges,
in dissipative multi-level atomic vapors
with linear, cubic and quintic susceptibilities present simultaneously.
The appearance of vortices results from a combined action of the number of topological charges, the
beam width of incidences, the diffusion effect, CQ nonlinearities, and the loss/gain in the medium,
all of them acting simultaneously.

We have formulated common rules for finding the number as well as the rotation direction of the induced vortices.
We also discover that the NTC of the vortex is conserved during propagation.
At last, some aspects of the liquid-like behavior of light in our system and potential experimental
investigations are also involved.

\section*{Acknowledgments}

This work was supported by CPSF (2012M521773),
the Qatar National Research Fund NPRP 09-462-1-074 project,
the 973 Program (2012CB921804), NSFC (61078002, 61078020, 11104214, 61108017, 11104216, 61205112),
RFDP (20110201110006, 20110201120005, 20100201120031),
and FRFCU (xjj2013089, 2012jdhz05, 2011jdhz07, xjj2011083, xjj2011084, xjj2012080).

\section*{Appendix}
\subsection*{The density matrix equations}
\label{appendix1}

Considering the time-dependent Schr\"odinger equation,
using a perturbation expansion and rotating wave approximation \cite{boyd_book},
we can obtain a series of density matrix equations as follows:
\begin{align}
\label{app1}  \frac{\partial }{{\partial t}}\rho _{00}^{(r)} = & - {\Gamma _{00}}\rho _{00}^{(r)} + i[ {G_1^*\rho _{10}^{(r - 1)} - {G_1}\rho _{01}^{(r - 1)}} ],\\
  \frac{\partial }{\partial t}\rho _{11}^{(r)} = & - \Gamma _{11} \rho_{11}^{(r)} + i[(G_1 \rho _{01}^{(r - 1)} - G_1^* \rho_{10}^{(r - 1)} ) \notag\\
\label{app2}  &+ G_2^\ast(\rho_{21}^{(r - 1)} - \rho_{12}^{(r - 1)} ) ],\\
  \frac{\partial }{{\partial t}}\rho _{10}^{(r)} =&  - ( {i{\Delta _1} + {\Gamma _{10}}} )\rho _{10}^{(r)} \notag\\
\label{app3}  &+ i[ {{G_1}( {\rho _{00}^{(r - 1)} - \rho _{11}^{(r - 1)}} ) + G_2^*\rho _{20}^{(r - 1)}}],\\
  \frac{\partial }{{\partial t}}\rho _{20}^{(r)} = & - [ {i( {{\Delta _1} + {\Delta _2}} ) + {\Gamma _{20}}} ]\rho _{20}^{(r)} \notag\\
\label{app4}  &+ i[ {G_2^*\rho _{10}^{(r - 1)} - G_1 \rho _{21}^{(r - 1)}} ],\\
  \frac{\partial }{\partial t}\rho _{21}^{(r)} = & - (i\Delta _2 + \Gamma _{21})\rho _{21}^{(r)} \notag\\
\label{app5}  &+ i[G_2(\rho _{11}^{(r - 1)} - \rho_{22}^{(r - 1)} ) - G_1^*\rho _{20}^{(r - 1)} ].
\end{align}

\subsection*{Derivation of the susceptibility}
\label{appendix2}

We display the detailed derivation process on $-\frac{\eta}{K^2}\frac{|G_1|^2}{d_1'}$, the first term in Eq. (\ref{chis3}),
by using the \textit{dressed perturbation chain method} \cite{fupanming_pra_2005,zhiqiang_pra_2008},
which involves the perturbation chain \cite{fupanming_pra_1995, yanpeng_pra_2001} and coupling equations together.

Firstly, a ground state particle $\rho_{00}^{(0)}$  absorbs a probe photon $p$  and transits to
$\rho _{G_2 \pm 0}^{(1)}$, the dressed state of $\rho_{10}^{(1)}$ ($\rho _{00}^{(0)}\xrightarrow{G_1}\rho _{G_2 \pm 0}^{(1)}$).
Under the weak field and steady state approximations,
the coupling equations can be obtained from Eqs. (\ref{app3}) and (\ref{app4})
\begin{equation*}
\begin{split}
  0 =&  - (i{\Delta _1} + {\Gamma _{10}}){\rho _{{G_2} \pm 0}} + i{G_1}{\rho _{00}} + i{G_2}{\rho _{20}},\\
  0 =&  - [i({\Delta _1} + {\Delta _2}) + {\Gamma _{20}}]{\rho _{20}} + iG_2^*{\rho _{{G_2} \pm 0}},
\end{split}
\end{equation*}
which gives
\begin{equation}\label{app6}
  \rho _{G_2 \pm 0}^{(1)} = \frac{{i{G_1}}}{{i{\Delta _1} + {\Gamma _{10}} + \frac{{|{G_2}{|^2}}}{{i({\Delta _1} + {\Delta _2}) + {\Gamma _{20}}}}}}\rho _{00}^{(0)}.
\end{equation}

Secondly, the particle absorb a pumping photon $p$  and transits to the dressed state  $\rho _{11}^{(2)}$ ($\rho _{G_2 \pm 0}^{(1)}\xrightarrow{G_1^\ast}\rho _{11}^{(2)}$).
From Eqs. (\ref{app2}) and (\ref{app5}) as well as the approximations used in the first step, we get the coupling equations
\begin{equation*}
\begin{split}
  {\Gamma _{11}}{\rho _{11}} = & i( - G_1^*{\rho _{{G_2} \pm 0}} + G_2^*{\rho _{21}}),\\
  (i{\Delta _2} + {\Gamma _{21}}){\rho _{21}} = & i{G_2}{\rho _{11}},
\end{split}
\end{equation*}
which gives
\begin{equation}\label{app7}
  \rho _{11}^{(2)} = \frac{iG_1^*} {\Gamma _{11} + \frac{|G_2|^2}{i\Delta_2 + \Gamma_{21}}} \rho _{G_2\pm0}^{(1)}.
\end{equation}

Thirdly, the stimulated atom transits back to state  ${\rho _{10}}$ and emits a pumping photon  $p^*$ ($\rho _{11}^{(2)}\xrightarrow{G_1^\ast}\rho _{G_2 \pm 0}^{(3)}$).
Similar to the first step, we get the coupling equations
\begin{equation*}
\begin{split}
  0 =&  - (i{\Delta _1} + {\Gamma _{10}}){\rho _{{G_2} \pm 0}} + i{G_1}{\rho _{11}} + i{G_2}{\rho _{20}},\\
  0 =&  - [i({\Delta _1} + {\Delta _2}) + {\Gamma _{20}}]{\rho _{20}} + iG_2^*{\rho _{{G_2} \pm 0}},
\end{split}
\end{equation*}
and the corresponding solution
\begin{equation}\label{app8}
  \rho _{G_2 \pm 0}^{(3)} = \frac{{i{G_1}}}{{i{\Delta _1} + {\Gamma _{10}} + \frac{{|{G_2}{|^2}}}{{i({\Delta _1} + {\Delta _2}) + {\Gamma _{20}}}}}}\rho _{11}^{(2)}.
\end{equation}

Combining Eqs. (\ref{app6})-(\ref{app8}),
and considering the assumption $\rho _{00}^{\left( 0 \right)} \simeq 1$ and the relations
$P = N \mu_{10} \rho_{G_2\pm0}^{(3)} = \epsilon_0 \chi^{(3)}|E_1|^2 {E_1}$,
we finally obtain the expression for the susceptibility
\begin{align}
\chi ^{(3)}|E_1|^2 =&  -\frac{iN\mu_{10}^2}{\hbar \epsilon _0}\frac{1}{\left(\Gamma_{10} + i\Delta_1 + \frac{|G_2|^2}{\Gamma_{20} + i(\Delta_1 + \Delta_2)} \right)^2} \notag\\
\label{app9} &\times \frac{|G_1|^2}{\Gamma_{11} + \frac{|G_2|^2}{\Gamma_{21} + i\Delta_2}}.
\end{align}

\bibliographystyle{unsrt}
\bibliography{pra_yiqi}

\end{document}